\documentclass[useAMS,usenatbib]{mn2e}
\usepackage[dvips]{graphics}

\title[Gas stripping by radiation drag]
{Gas stripping by radiation drag from an interstellar cloud }
\author[Jun'ichi Sato, Masayuki Umemura, Keisuke Sawada 
and Shingo Matsuyama]
{Jun'ichi Sato$^{1}$
\thanks{E-mail:junichi@rccp.tsukuba.ac.jp (JS)},
Masayuki Umemura$^{1}$
\thanks{E-mail:umemura@rccp.tsukuba.ac.jp (MU)}, 
Keisuke Sawada$^{2}$
\thanks{E-mail:sawada@cfd.mech.tohoku.ac.jp (KS)} 
and Shingo Matsuyama$^{3}$
\thanks{E-mail:smatsu@chofu.jaxa.jp (SM)}
\\
$^{1}$Center for Computational Science, University of Tsukuba, 
Tsukuba 305-8577, Japan
\\
$^{2}$Department of Aeronautics and Space Engineering, Tohoku University, 
Sendai 980-8579, Japan
\\
$^{3}$Institute of Space Technology and Aeronoutics, JAXA, 
Chofu 182-8522, Japan}
\begin{document}

\date{in original form 2004 January 15}

\pagerange{\pageref{firstpage}--\pageref{lastpage}} \pubyear{2004}

\maketitle

\label{firstpage}

\begin{abstract}
We perform two-dimensional hydrodynamic simulation 
on the gas stripping by radiation drag from an interstellar cloud 
moving in uniform radiation fields. 
To properly include relativistic radiation drag, the radiation hydrodynamic 
equation is solved with taking into account the dilution of radiation 
fields by optical depth of the cloud.
As a result, it is found that the optically-thin surface layers are 
effectively stripped by radiation drag from an optically-thick gas cloud,
and simultaneously stripped gas loses momentum.
The momentum loss time-scale is found to be 
on the order of $10^{8}$ years under intensive radiation fields 
which are expected in the early phase of galaxy evolution.
The present results show that the radiation drag is an effective 
mechanism to extract angular momentum from interstellar medium
and allows it to accrete onto the galactic center.
The mass accretion driven by radiation drag may lead to 
the formation of a central supermassive black hole.
\end{abstract}


\section{Introduction}\label{section-1}
If the matter moves in radiation fields, it absorbs the radiation anisotropically
in comoving frame and then reemits the radiation isotropically, so that
the gas is subject to drag force in proportion to 
$vE$, where $v$ is the peculiar velocity and $E$ is the radiation energy
density \citep{MM84}.
This effect is known as the radiation drag.
A famous example of radiation drag is the Poynting-Robertson 
effect on a dust grain in the Solar System 
\citep{Poynting03,Robertson37}, where, 
due to the angular momentum loss by the radiation drag,
a grain spirals into the Sun.

The importance of radiation drag has been 
considered for other various issues in astrophysics.
In the epochs shortly after the recombination,
the cosmic background radiation can exert strong drag force 
through electron scattering, which is called the Compton drag.
\citet{Loeb93} considered the possibility that the Compton drag
could extract angular momentum from an early formed
object and build up a massive black hole. In this context,
\citet{ULT93} performed numerical simulations,
using a three-dimensional hydrodynamics-dark-matter code, 
and found that the angular momentum extraction actually 
takes place efficiently at redshifts of $z \ga 300$, leading
to the formation of massive black holes.

The radiation drag induced mass accretion is also considered 
in a rotating gas disk immersed in the external radiation fields.
Cosmological accretion disks embedded in background
radiation are extensively studied by \citet{FU94}, \citet{TFU94}, 
\citet{UF94}, \citet{TU97}, and \citet{MTU98}. 
In relation to quasi-periodic phenomena in X-ray stars, 
\citet{FLM89}, \citet{Lamb89, Lamb91} and \citet{ML93} 
examined the role of radiation (Compton) 
drag exerted by the radiation fields of a central neutron star on the 
disk gas in the binary regime. 
\citet{MMH94} proposed siphon processes in cataclysmic variables, 
where the evaporating gas from the disk surface infalls toward 
a central white dwarf because of radiation drag. 
\citet{FU95} investigated a steady accretion driven by the radiation
of a central luminous object. 
Also, in active galactic nuclei (AGNs), the radiation drag may play
an important role in some situations.
\citet{UFM97} proposed 
the starburst-induced fueling to an AGN ("radiative avalanche"),
to relate the circumnuclear starburst with AGN event. 
This model has been extended
to more realistic versions \citep{FUM97,UFM98,Ohsuga+99}.
Commonly, it was found that the surface layer of rotating gas irradiated 
by intensive starlight can lose the angular momentum 
by radiation drag, which results in an avalanche of the layer 
as an inevitable consequence. 

The model of mass accretion induced by uniform background
radiation is revitalized in galactic bulges,
because bulge stars produce nearly uniform radiation fields.
This can provide a possible way to cause the formation of 
a supermassive black hole (SMBH) in a bulge.
The recent observations have suggested that the mass of 
a SMBH tightly correlates with the mass of the host bulge 
or the velocity dispersion of bulge stars
\citep{KR95,Miyoshi+95,Laor98,Magorrian+98,Richstone+98,Ho99, 
Wandel99,KH00,Gebhardt+00a,Gebhardt+00b,FM00,MFT00,Nelson00, 
Salucci+00,Ferrarese+01,MD01,MF01a,MF01b,Sarzi+01,MD02,Wandel02}. 
All of these correlations imply that the formation of a SMBH is 
likely to be physically linked to the formation of a galactic 
bulge that harbors a SMBH. 
\citet{Umemura01} considered the accretion of dusty gas
driven via the radiation drag by bulge stars and constructed
a quasar formation model.
\citet{KU02} quantitatively clarified by
radiative transfer calculations that the inhomogenity
of dusty interstellar medium controls the efficiency
of radiation drag, and found that, 
if surface gas layers are stripped effectively by
the radiation drag from optically-thick dusty gas clouds
and lose angular momentum, 
a sufficient amount of gas can accrete to 
form a SMBH in realistic situations.
The model of radiation drag-induced formation of 
SMBH predicts the SMBH-to-bulge mass relation
to be 0.3-0.5$\varepsilon$, where $\varepsilon$
is the fraction of $mc^2$ liberated by hydrogen burning in stars.
However, the processes of gas stripping by the radiation drag 
from an optically-thick cloud have not been elucidated in detail.
The study requires a sophisticated numerical simulation
with relativistic radiation hydrodynamics.
In this paper, we perform a radiation hydrodynamic simulation 
with generalized coordinates
for a gas cloud moving in radiation fields.
Through this simulation, we attempt to demonstrate
the stripping efficiency and the time-scale of angular momentum loss. 

The paper is organized as follows. 
In Section \ref{section-2}, we present the model used in this work. 
In Section \ref{section-3}, we describe the numerical treatment. 
In Section \ref{section-4}, we show the numerical results
on the flow patterns and then estimate the 
time-scale for momentum loss of the gas. 
Section \ref{section-5} is devoted to conclusions. 

\section[]{MODEL}\label{section-2}
We suppose a simple situation that a spherical cold gas cloud 
moves in uniform radiation fields with a constant velocity.
The gas cloud is assumed to have a uniform density distribution
and to be in pressure balance with hot rarefied ambient gas.
If there is relatively large velocity difference between the gas cloud 
and the ambient matter, the cloud surface layers are expected to be
stripped by the ram pressure. The ram-pressure stripping
itself does not extract angular momentum from the total system
including the cloud and ambient matter. But, if the system is embedded 
in radiation fields, the optically-thin parts of the stripped gas can
be subject to the radiation drag and lose angular momentum.
This might provides a simple way to extract angular momentum
from the interstellar medium. From a realistic point of view, however,
the gas motion does not seem so simple. Recent high resolution
simulations on multiple supernova (SN) explosions in a star-forming 
galaxy \citep{MUF04} have reveal that
dense shells driven by SN hot bubbles undergo thermal instability
and resultant cold clouds move together with ambient hot gas. 
Here, we consider two cases; one is a comoving case
in which the stripping takes place solely by the radiation drag,
and the other is a shear-flow case in which the radiation drag is 
partially coupled with the ram-pressure stripping.

The cloud temperature, mass, and radius are respectively assumed 
to be $T_{c}=10^{2} \mbox{K}$, 
$M_{c}=3.0 \times 10^{5} \mbox{M}_{\odot}$, and $R_{c}=10 \mbox{pc}$, 
while the temperature of hot ambient matter is
$T_{h}=10^{7} \mbox{K}$.
We postulate the peculiar velocity $ v_{p}$ of the cloud to be of the order of
the virial velocity for a spheroidal galaxy 
with $M_{\mbox{tot}}=10^{11} \mbox{M}_{\odot}$
and $R_{g}=1 \mbox{kpc}$:
\begin{equation}
v_{\rm vir}=6.67 \times 10^{2} \mbox{km s}^{-1}
\left(
\frac{M_{\mbox{tot}}}{10^{11} \mbox{M}_{\odot}}
\right)^{\frac{1}{2}}
\left(
\frac{R_{g}}{1 \mbox{kpc}}
\right)^{-\frac{1}{2}}.
\end{equation}
As for the radiation fields, the radiation energy density
is estimated by the assumed total luminosity $L_{*}$ 
and galaxy size to be 
\begin{equation}
E= {L_{*} \over 4\pi R_{g}^2 c} =
3.45 \times 10^{-2} \mbox{erg cm}^{-3}
\left( \frac{L_{*}}{3 \times 10^{12} \mbox{L}_{\odot}} \right)
\left( \frac{R_{g}}{1 \mbox{kpc}} \right)^{-2}. 
\end{equation}
In this paper we assume the total luminosity to be 
$L_{*}= 3 \times 10^{12} \mbox{L}_{\odot}$. 

\section{METHOD OF CALCULATIONS}\label{section-3}

\begin{figure*}
\vspace{7cm}
\caption{The two-dimensional generalized curvilinear coordinate used here. 
         The number of grid points is $101 \times 101$. 
         The surface of the cold gas is represented by a dash-dotted line.}
\includegraphics{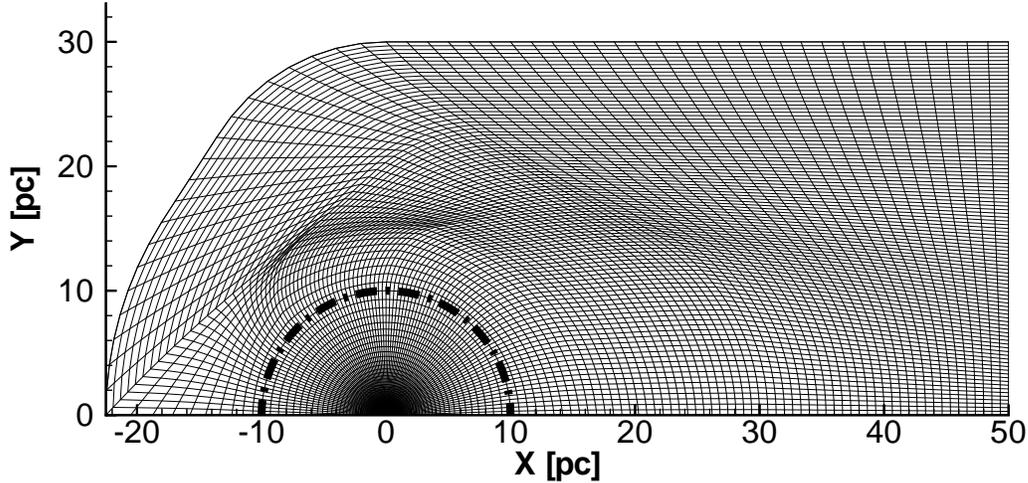}
\label{fig-1}
\end{figure*}

We use the two-dimensional 
generalized curvilinear coordinates as shown in Fig. \ref{fig-1}. 
The number of grid points is $101 \times 101$. 
Such coordinates allow us to take accurate
boundary conditions for a spherical gas cloud
and also to treat appropriately the flow of stripped gas 
on the right-hand side.
The scale of computational domain is about $50 \mbox{pc}$. 
The gas cloud moves at a velocity $v_{p}$ along the $x$-axis 
in the radiation fields. Here, to prevent the cloud from
going out of the boundary with $v_{p}$, we make
the Galilean transformation by $v_{p}$ so that the cloud should
stay at the origin if neither drag force nor ram pressure is exerted.
In these coodinates,
the surface of the cold gas is represented by a dash-dotted line 
in Fig. \ref{fig-1}. The other regions are filled with the hot gas. 

The two-dimensional Euler equations including the relativistic
radiation drag term can be written in the conservation form as 
\begin{equation}
\frac{\partial \tilde{Q}}{\partial t}+
\frac{\partial \tilde{F}}{\partial x}+
\frac{\partial \tilde{G}}{\partial y}+
\tilde{H}=0, 
\label{eq-1}
\end{equation}
where $\tilde{Q}$, $\tilde{F}$, $\tilde{G}$ and $\tilde{H}$ 
are respectively the following vectors, 
\begin{eqnarray}
&&
\tilde{Q}=
\left(
\begin{array}{c}
\rho \\
\rho u \\
\rho v \\
e
\end{array}
\right)
,\quad
\tilde{F}=
\left(
\begin{array}{c}
\rho u \\
\rho u^{2}+p \\
\rho uv \\
(e+p)u
\end{array}
\right)
,
\nonumber \\
&&
\tilde{G}=
\left(
\begin{array}{c}
\rho v \\
\rho uv \\
\rho v^{2}+p \\
(e+p)v
\end{array}
\right)
,
\nonumber \\
&&
\tilde{H}=
\left(
\begin{array}{c}
0 \\
\rho \chi E\left(u-v_{p}\right)/c\\
\rho \chi E v/c\\
0
\end{array}
\right),
\label{eq-2}
\end{eqnarray}
where $\rho$ is the density of the gas, $u$ and $v$ are 
respectively the $x$- and $y$- components of velocity,
$c$ is the speed of light, and
$\chi$ is the mass extinction coefficient due to dust opacity.
In this work, we set
$\chi =100 \mbox{cm}^{2} \mbox{g}^{-1}$
by assumig the Solar abundance and $0.1\mu$m dust grains
(e.g. see Umemura et al. 1998).
The pressure $p$ and the total energy per unit volume $e$ 
are related by the equation of state, 
\begin{equation}
p=\left(\gamma-1\right)
\left\{e-\frac{\rho}{2}\left(u^{2}+v^{2}\right)\right\},
\label{eq-3}
\end{equation}
where $\gamma$ is the ratio of specific heats of the gas, 
which is chosen as $\gamma=5/3$. 
The radiation drag effect is taken into account 
by adding the term of $\tilde{H}$. 

To discretise the governing equation, eq (\ref{eq-1}),
the finite volume-approach is used, where
the AUSM-DV scheme is employed to obtain the numerical 
convective flux \citep{WL94}. The time integration of
a set of equations is performed implicitly 
by a matrix free LU-SGS method \citep{Nakahashi+99}. 

\begin{figure}
\vspace{7cm}
\caption{The distribution of the optical depth in the vicinity of 
         the cold gas at initial time. 
         The optical depth at each point is evaluated by integrating 
         the opacity along with a line drawn from the outer boundary to 
         the coordinate origin. 
         The surface of the cold gas is represented by a dash-dotted line.}
\includegraphics{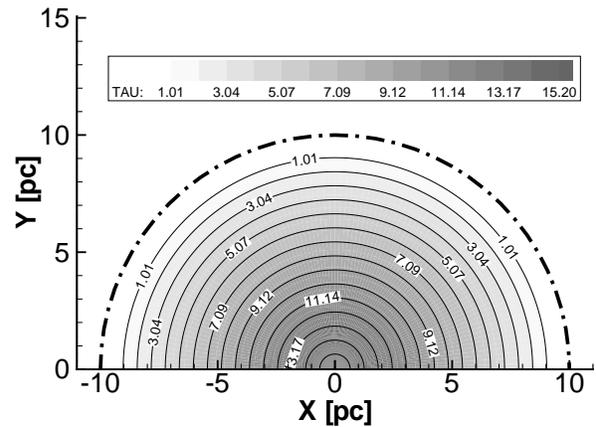}
 \label{fig-2}
\end{figure}

The optical depth at each point is evaluated by integrating 
the opacity along with a line drawn from the outer boundary to 
the coordinate origin. 
The distribution of the optical depth in the cloud
at initial time is shown in Fig. \ref{fig-2}. 
The optical depth at the center of cloud
reaches above $15$ and the thickness of 
the optically-thin surface layer is $\sim 10 \%$ of the cloud radius. 
We involve the radiation drag term if the optical depth is
lower than unity, because no drag force works in optically-thick 
medium \citep{TU97}. 

\begin{figure}
\vspace{17cm}
\caption{The optical depth distributions on several snapshots of
         the dynamical evolution. The dot-dashed semicircle shows 
         the initial position of the cloud. 
         The light gray or charcoal gray regions show 
         the domain where the optical depth is 
         from $0.1$ to $1$ or above $1$, 
         respectively. The gas in light gray is subject to 
         the radiation drag, while that in charcoal gray regions is 
         impervious to the radiation drag.}
\includegraphics{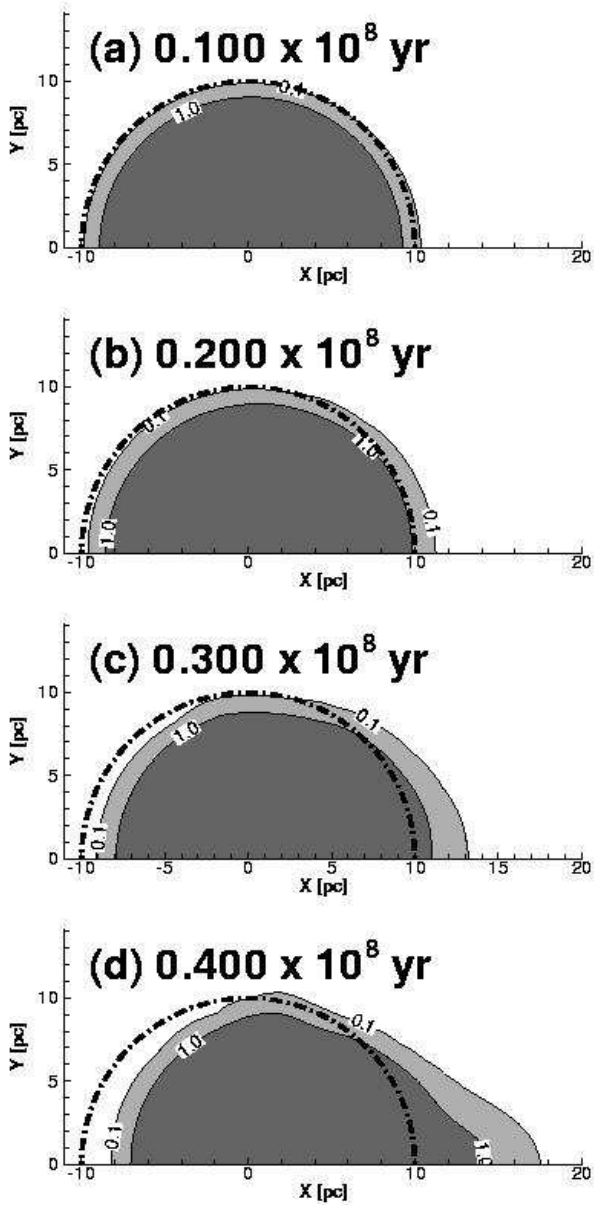}
\label{fig-3}
\end{figure}

In order to demonstrate the validity of the present method 
for the optical depth estimation, we show the 
optical depth distributions on several snapshots of
the dynamical evolution in Fig. \ref{fig-3}. 
The dot-dashed semicircle shows the initial position of the cloud. 
The light gray or charcoal gray regions show 
the domain where the optical depth is from $0.1$ to $1$ or above $1$, 
respectively. The optically-thin regions are properly detected 
with tracing the dynamical evolution. 
The gas in charcoal gray regions is impervious to
the radiation drag. 
However, there is limitation for applying this method.
If the whole region of the cold gas deviates 
from the coordinate origin, the estimate of optical depth 
would not be correct. 
Threfore, at this time, the estimate of the radiation drag 
is not always accurate. 
In Section \ref{section-4} we will be back to this point. 

\section{RESULTS}\label{section-4}

\subsection{Comoving Case}

\begin{figure}
\vspace{17cm}
\caption{The time variation of density distributions in a comoving case. 
         (a), (b), (c) and (d) show these at time, 
         $0.100 \times 10^{8}$,
         $0.388 \times 10^{8}$,
         $0.504 \times 10^{8}$, and 
         $1.252 \times 10^{8} \mbox{yr}$, respectively. 
         The density range is from $3.5 \times 10^{-23}$ to 
         $5.3 \times 10^{-21} \mbox{g cm}^{-3}$. 
         Each density range 
         is divided by $21$ lines with an equal increment. }
\includegraphics{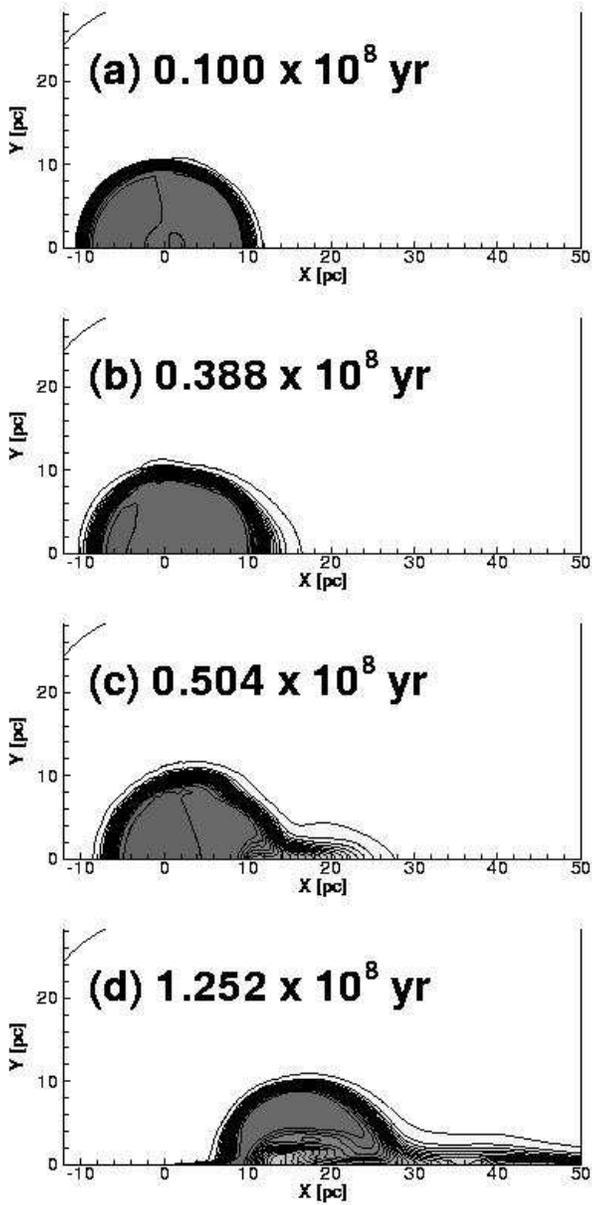}
\label{fig-4}
\end{figure}

\begin{figure}
\vspace{17cm}
\caption{The time variation of vector fields for the optically-thin cold gas 
         in a comoving case. 
         (a), (b), (c) and (d) show the flow patterns at the same stages as 
         Fig. \ref{fig-4}, respectively. 
	 The density contours are also displayed. 
         The density range and the number of contour lines are 
         also the same as Fig. \ref{fig-4}. }
\includegraphics{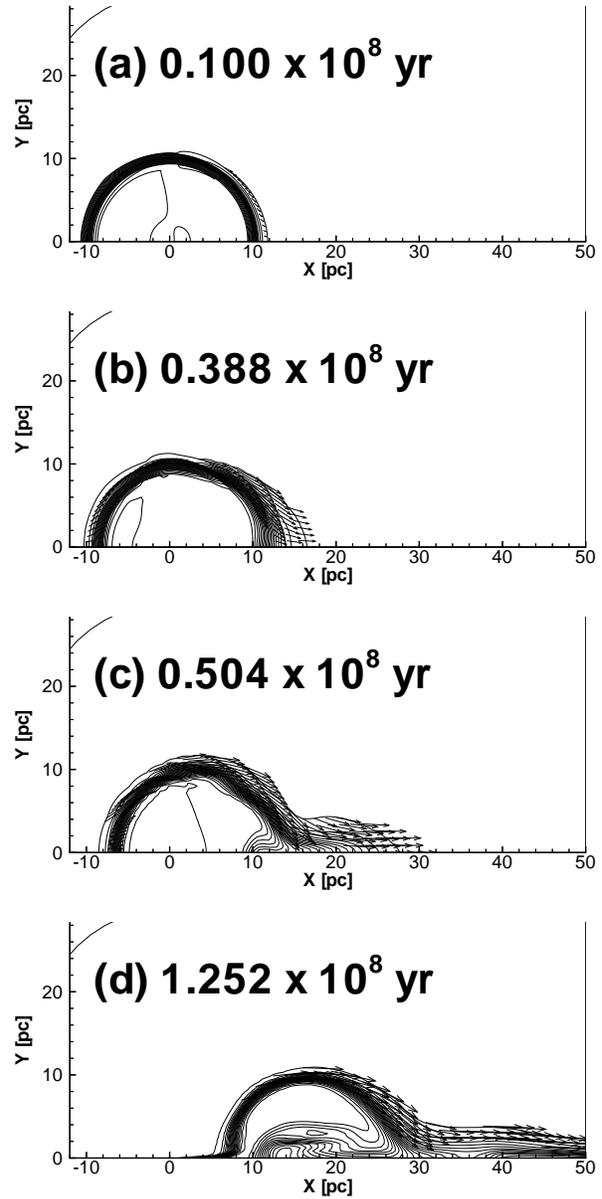}
\label{fig-5}
\end{figure}

The stripping purely by the radiation drag is demonstrated 
in a comoving case, where no relative velocity is present 
between the gas cloud and the ambient matter. 
The time variation of density distributions
is depicted by four snapshots in Fig. \ref{fig-4},
and that of velocity fields is shown only for 
optically-thin cold gas in Fig. \ref{fig-5}, where
the density contours are also displayed.
We can see in Fig. \ref{fig-4} that the surface layers of 
the gas cloud are stripped by radiation drag and drifted 
to the right-hand side. 
Since we are observing the phenomena 
in the coordinates moving to the left at the velocity $v_{p}$,  
the gas drifted to the right is actually 
left behind by the radiation drag. 
It is noted that this stripping is obviously different from 
the ram-pressure stripping in that it occurs also 
behind the cloud in the downstream. 
The left-hand side of the optically-thin surface layer 
pushes the whole region to the right by the radiation drag. 

The velocity fields in Fig. \ref{fig-5} show that
the optically-thin surface layers move round the surface of the cloud. 
A part of the gas reaching the $x$-axis is accelerated 
toward the right direction along the $x$-axis. 
This means that the stripped gas slows down by the radiation drag.

In the later phase ($t>10^8$yr),
the stripped gas is trailed by the radiation drag,
because the cold gas is confined by the ambient hot gas. 
When the stripped gas reaches the $x$-axis, it 
splits into the left and right under the influence of external pressure.
The radiation drag is not exerted on the gas splitting into the left 
owing to the large optical depth and therefore left there 
in the present Galilean transformed coordinates, 
while the radiation drag works to decelerate
the optically-thin gas splitting to the right and 
eventually produces the gas flows to the right. 
As mentioned before, when the whole region of the cloud deviates 
from the origin, the present estimate of the optical depth 
in the left-hand side of the cloud becomes inaccurate. 
However, as far as the gas drifted to the right is concerned, 
the optical depth is still properly estimated. Hence,
the trailing flow is thought to be real.

\subsection{Shear-Flow Case}

\begin{figure}
\vspace{17cm}
\caption{The time variation of density distributions in a shear-flow case. 
         (a), (b), (c) and (d) show these at time, 
         $0.100 \times 10^{8}$,
         $0.380 \times 10^{8}$,
         $0.508 \times 10^{8}$, and 
         $0.862 \times 10^{8} \mbox{yr}$, respectively. 
         The density range and the number of contour lines are 
         the same as Fig. \ref{fig-1}. }
\includegraphics{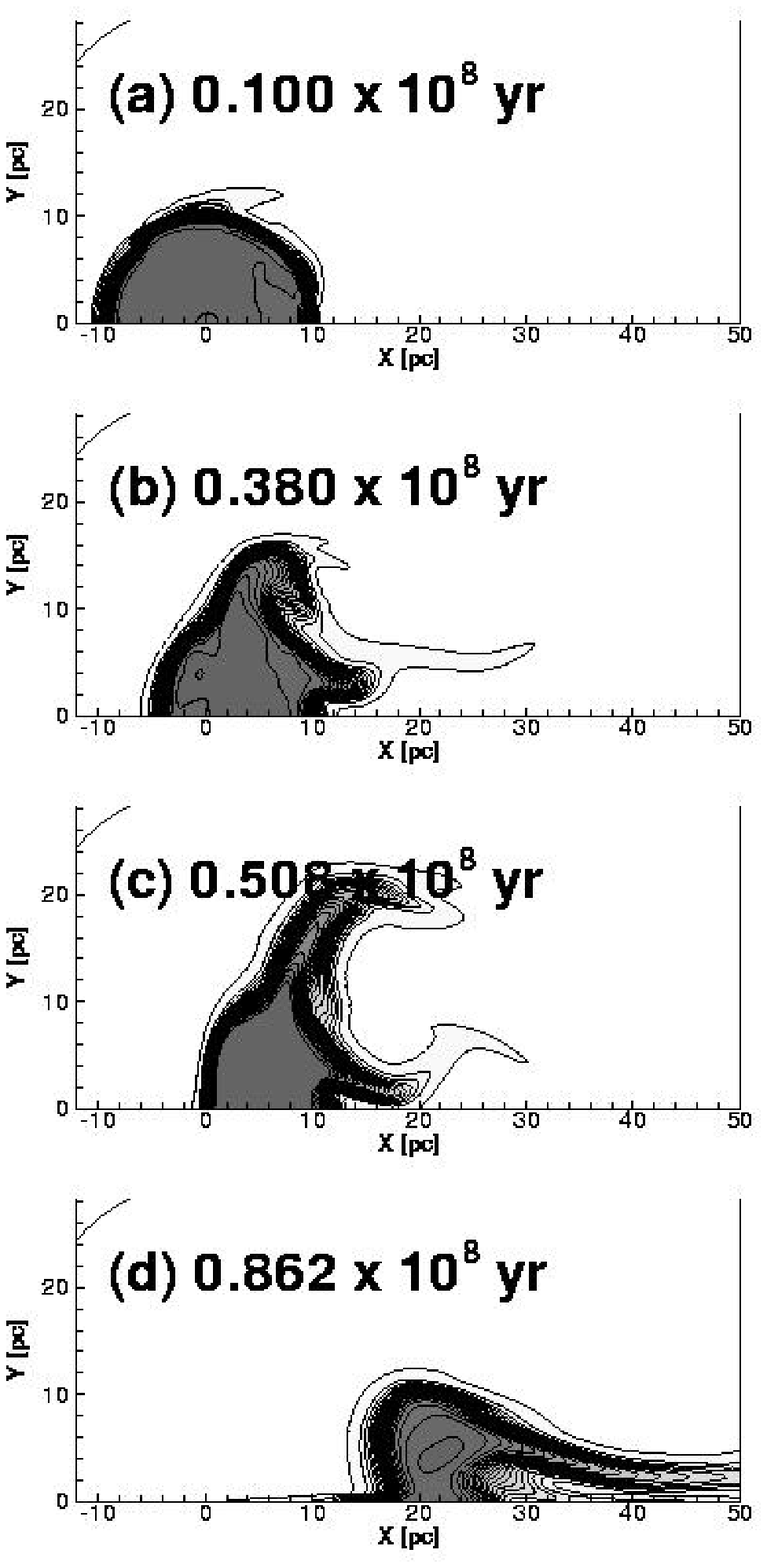}
\label{fig-6}
\end{figure}

\begin{figure}
\vspace{17cm}
\caption{The time variation of vector fields for the optically-thin cold gas 
         in a shear-flow case. 
         (a), (b), (c) and (d) show the flow patterns at the same stages as 
         Fig. \ref{fig-6}, respectively. 
	 The density contours are also displayed. 
         The density range and the number of contour lines are 
         the same as Fig. \ref{fig-4}. }
\includegraphics{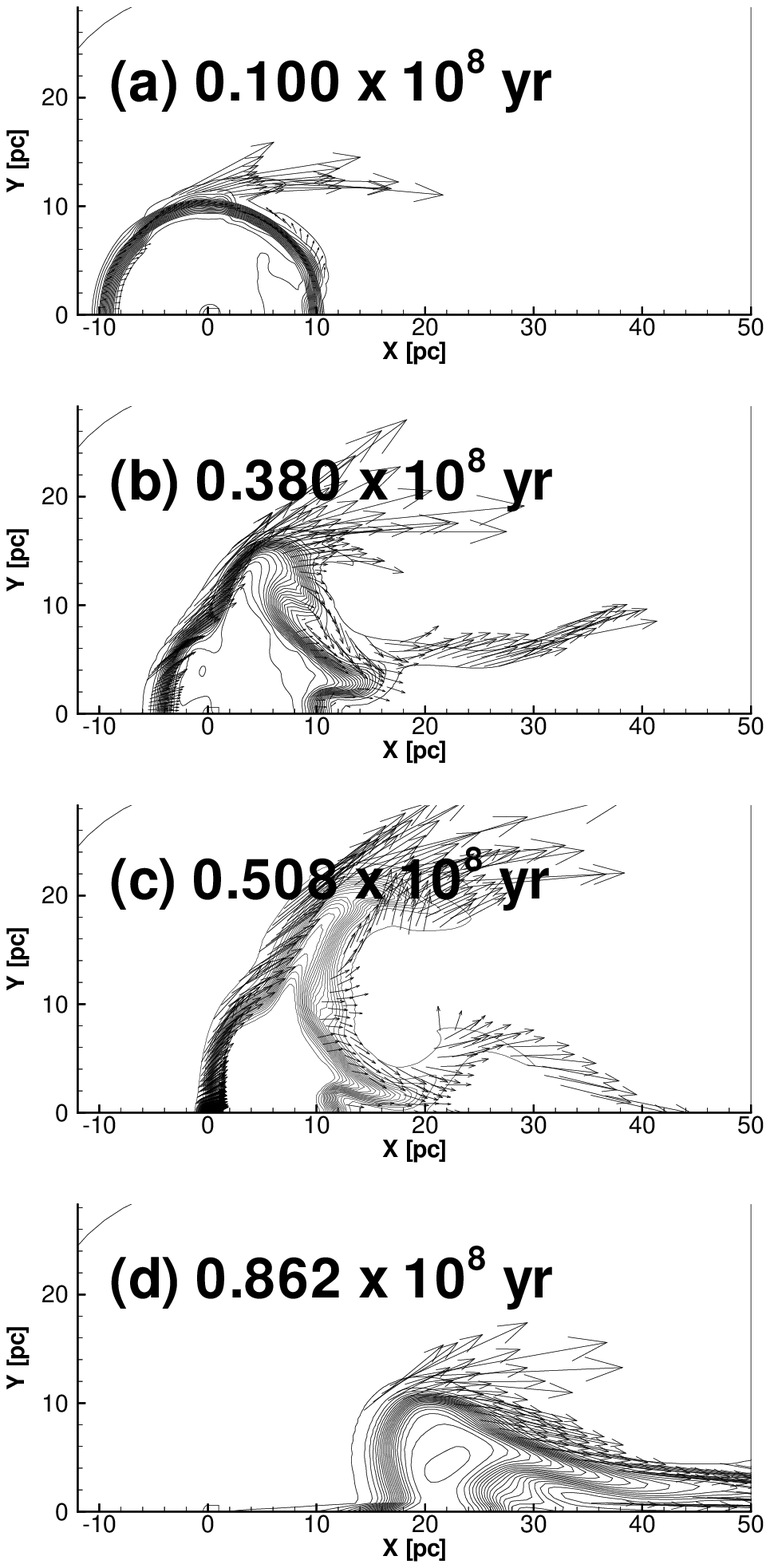}
\label{fig-7}
\end{figure}

Here, we show the results for a shear-flow case in which
the velocity difference between the gas cloud and the ambient matter
is assumed to be $0.1 v_{p}$, 
corresponding to $0.13$ in Mach number. 
The numerical results are shown in Fig. \ref{fig-6} and Fig. \ref{fig-7}
respectively for the density distributions and the velocity fields
for the optically-thin cold gas.

The snapshots of the density contours in Fig. \ref{fig-6}
show that the deformation of the gas cloud grows with time. 
In this case, the ram-pressure stripping occurs simultaneously 
with the stripping driven by the radiation drag. 
However, as shown in Fig. \ref{fig-6} (d), 
the gas drifted to the right eventually flows along with the $x$-axis. 
This flow pattern is similar to that in the case without 
the ram-pressure induced deformation. 

The snapshots of the velocity fields
show that the rightward velocities 
of the optically-thin surface layers increase with the deformation. 
Our numerical method allows us to trace the optically-thin surface 
even in such strong deformation of the cold gas. 
The optically-thin gas is subject to the radiation drag and
makes a trailing flow, similar to the case without 
the ram pressure. 

\subsection{Momentum Loss}

\begin{figure}
\vspace{8cm}
\caption{The time variation of momentum in units of the initial momentum. 
         The solid line is the evolution in a comoving case, 
         while the broken line is in a shear-flow case.  
         A dotted line shows the level of $1/e$ reduction. }
\includegraphics{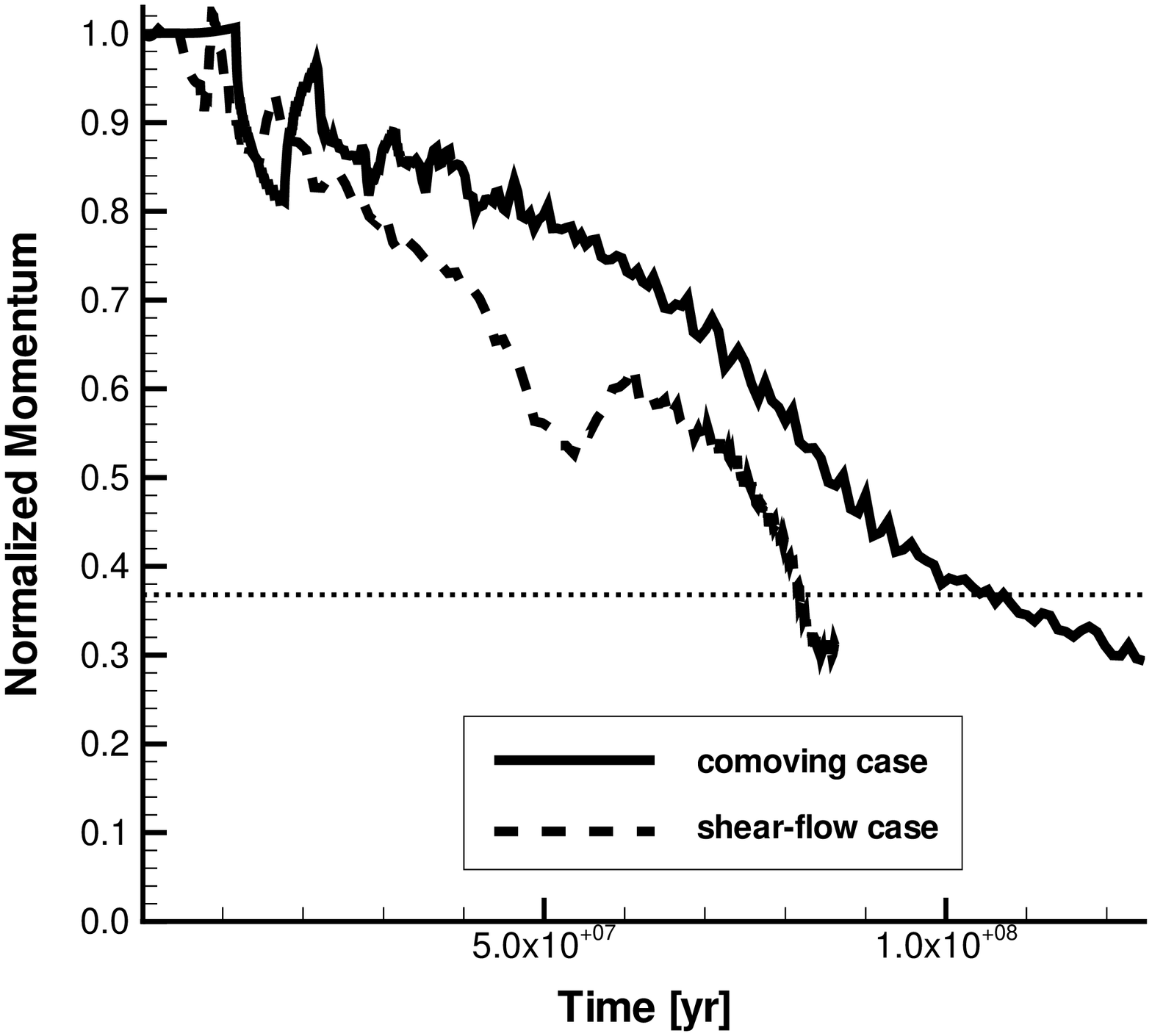}
\label{fig-8}
\end{figure}

The radiation drag works on both radial and azimuthal components of
gas velocity, in proportion to each component. 
In a pure rotating system, the reduction only for
azimuthal velocity leads to the angular momentum loss.
But, in a chaotic system like a bulge under starbursts,
the reduction of random velocities means, on average, the decrease
of an azimuthal component of velocity and therefore angular momentum. 
In other words, the gas even in non-radial motion can fall onto 
the center if the momentum of the gas is strongly reduced. 
Hence, here we consider the momentum loss in the
present simulations.
The momentum of optically-thin cold gas in the 
$x$-direction, $Q_{x}$, is defined as 
\begin{equation}
Q_{x} \equiv \int_{\tau \leq 1} 
\rho \mid u-v_{p} \mid dV.
\end{equation}
In Fig. \ref{fig-8}, the time variation of momentum is shown
in units of the initial momentum.
The solid line is the evolution in a comoving case, while
the broken line is in a shear-flow case. 
A dotted line shows the level of $1/e$ reduction.

From this result, 
we can evaluate the momentum loss time-scale, which
is defined by the $1/e$ reduction of momentum.
The time-scale for a comoving case 
is $t=1.04 \times 10^{8} \mbox{yr}$, and
that for a shear-flow case is $t=0.8 \times 10^{8} \mbox{yr}$.
In the shear-flow case, the surface gas of the cloud is stripped
also by the ram pressure and is rarefied. Resultantly, 
the fraction of optically-thin gas increases compared to
the pure radiation-drag case, so that the momentum is 
extracted more efficiently than the comoving case.

The time-scale of angular momentum extraction is
analytically estimated by \citet{UFM98} to give
\begin{eqnarray}
t_{\gamma} &\simeq& 2.4 \times 10^{8} \mbox{yr}
\left(
\frac{L_{*}}{3 \times 10^{12}\mbox{L}_{\odot}}
\right)^{-1}
\left(
\frac{R_g}{\mbox{1 kpc}}
\right)^{2}\nonumber\\
&&\times 
\left(
\frac{f_{\mbox{dg}}}{10^{-1}}
\right)^{-2}
\left(
\frac{a_{\mbox{d}}}{0.1 \mu \mbox{m}}
\right)
\left(
\frac{\rho_{\mbox{s}}}{\mbox{g cm}^{-3}}
\right), 
\end{eqnarray}
where $f_{\mbox{dg}}$, $a_{\mbox{d}}$, and $\rho_{\mbox{s}}$ 
are the dust-to-gas mass ratio, the grain radius, and 
the density of solid material within the grain, respectively. 
It is found that the numerically obtained time-scale is
in a good agreement with the analytic estimate,
although the numerical one is slightly shorter.

\section{CONCLUSIONS AND DISCUSSION}\label{section-5}
We have performed the radiation-hydrodynamic simulation 
for a gas cloud moving at a constant velocity 
in uniform radiation fields, with focusing
attention on the radiation drag as well as the coupling effect
with the ram pressure. 
The obtained results are summarized as follows:

\begin{description}
\item[(1)] 
In a comoving case, the stripping occurs purely through
the radiation drag. All the optically-thin surface layers
are stripped from an optically-thick gas cloud,
in contrast to the pure ram-pressure stripping. 
Finally, a trailing flow behind the cloud forms.
\item[(2)] 
In a shear-flow case, the radiation drag-induced stripping 
occurs simultaneously with the ram-pressure stripping.
But, the eventual flow pattern for optically-thin gas is similar 
to the pure radiation drag case. 
\item[(3)] The momentum loss time-scale 
for the stripped gas is on the order of $10^{8}$ years 
under intensive radiation fields expected 
in an early evolutionary phase of a massive bulge. 
The time-scale is shorter for a shear-flow case with the aid of
the ram pressure.
The present time-scale is basically in a good agreement 
with an analytic estimation for angular momentum extraction 
time-scale. 
\end{description}

It is demonstrated that the radiation drag by bulge stars 
is an effective mechanism to extract angular momentum 
from interstellar cloud in a galactic bulge system. 
This mechanism allows the mass accretion onto the galactic center,
possibly leading to the formation of a supermassive black hole 
\citep{Umemura01, KU02, KUM03, KU04}.

In the present work, we have ignored the effect of self-gravity.
The self-gravity is likely to be important for a cloud 
with $M_{c}=3.0 \times 10^{5} \mbox{M}_{\odot}$ 
assumed in the present paper.
Thus, an assumption of a uniform cloud may not be
realistic, and instead the density distribution of a cold gas 
in gravitational equilibrium should be considered.
In this case, the outside of a gas cloud could be rarefied 
and the domain of the optically-thin layers increases. 
Consequently, the efficiency of the radiation drag
might be enhanced. 
To investigate the details of the stripping and momentum loss
in such a case will be significant in the future work. 

\section*{Acknowledgements}
We thank N. Kawakatu for many useful comments. 
The analysis has been made with computational facilities 
at Center for Computational Science in University of Tsukuba. 
MU acknowledges Research Grant from Japan Society for the Promotion of
Science (15340060).

\end{document}